\documentclass[12pt]{article}

\usepackage{amsmath}
\usepackage{amssymb}
\usepackage{amstext}
\usepackage{amsopn}
\usepackage{amsfonts}
\usepackage{amsxtra}
\usepackage[english]{babel}
\usepackage{graphicx}
\usepackage{float}
\usepackage{bm}
\usepackage{multirow}
\usepackage{hyperref}
\usepackage{lineno}
\usepackage{color}
\usepackage[superscript]{cite}
\usepackage{times}
\usepackage{fullpage}

\makeatletter
\renewcommand\@biblabel[1]{#1.}
\makeatother

\newcommand{\icm}{cm$^{-1}$}% % cm-1

\begin{document}

%%%%%%%%%%%%%%%%%%%%%%%%%%%%%%%%%%%%%%%%%%%%%%%%%%%%%%%%%%%%%%%%%%%%%
%%
%%  Title
%%

\begin{trivlist}
  \item[] {\Large\textsf{\textbf{Temperature-tunable Fano resonance induced by strong coupling between Weyl fermions and phonons in TaAs}}}
\end{trivlist}

%%%%%%%%%%%%%%%%%%%%%%%%%%%%%%%%%%%%%%%%%%%%%%%%%%%%%%%%%%%%%%%%%%%%%%%%%%%%%
%%
%%  Author list
%%
%%  Put the author list here
%%

\begin{trivlist}
  \item[] B. Xu$^{1,2,\dagger}$, Y. M. Dai$^{3,\dagger}$, L. X. Zhao$^{1}$, K. Wang$^{1}$, R. Yang$^{1}$, W. Zhang$^{1}$, J. Y. Liu$^{1}$, H. Xiao$^{2}$, G. F. Chen$^{1,4}$, S. A. Trugman$^{3,5}$, J.-X. Zhu$^{3,5}$, A. J. Taylor$^{6}$, D. A. Yarotski$^{3}$, R. P. Prasankumar$^{3,\ast}$, and X. G. Qiu$^{1,4,\ast}$

%%%%%%%%%%%%%%%%%%%%%%%%%%%%%%%%%%%%%%%%%%%%%%%%%%%%%%%%%%%%%%%%%%%%%%%%%%%%%%%%%%
%%
%%  affiliations
%%

 \item[] $^{1}$\emph{Beijing National Laboratory for Condensed Matter Physics, Institute of Physics, Chinese Academy of Sciences, P.O. Box 603, Beijing 100190, China}
 \item[] $^{2}$\emph{Center for High Pressure Science and Technology Advanced Research, Beijing 100094, China}
 \item[] $^{3}$\emph{Center for Integrated Nanotechnologies, Los Alamos National Laboratory, Los Alamos, New Mexico 87545, USA}
 \item[] $^{4}$\emph{Collaborative Innovation Center of Quantum Matter, Beijing 100190, China}
 \item[] $^{5}$\emph{Theoretical Division, Los Alamos National Laboratory, Los Alamos, New Mexico 87545, USA}
 \item[] $^{6}$\emph{Associate Directorate for Chemistry, Life and Earth Sciences, Los Alamos National Laboratory, Los Alamos, New Mexico 87545, USA}

%%%%%%%%%%%%%%%%%%%%%%%%%%%%%%%%%%%%%%%%%%%%%%%%%%%%%%%%%%%%%%%%%%
%%
%%  Correspondence, contact information
%%

  \vspace{2mm}
  \item[] $^{\dagger}$These authors contributed equally to this work.
  \item[] $^{\ast}$e-mail: rpprasan@lanl.gov; xgqiu@iphy.ac.cn
  \vspace{4mm}
\end{trivlist}

%\newpage
%%%%%%%%%%%%%%%%%%%%%%%%%%%%%%%%%%%%%%%%%%%%%%%%%%%%%%%%%%%%%%%%%%%%%%%%%%%%%%
%%
%%  Abstract
%%

\boldmath
\begin{trivlist}
\item[] {\bf Strong coupling between discrete phonon and continuous electron-hole pair excitations can give rise to a pronounced asymmetry in the phonon line shape, known as the Fano resonance\cite{Fano1961PR}. This effect has been observed in a variety of systems, such as stripe-phase nickelates\cite{Coslovich2013NC}, graphene\cite{Kuzmenko2009PRL,Tang2010NN,Li2012PRL} and high-$T_{c}$ superconductors\cite{Xu2015PRB}. Here, we reveal explicit evidence for strong coupling between an infrared-active $A_1$ phonon and electronic transitions near the Weyl points (Weyl fermions) through the observation of a Fano resonance in the recently discovered Weyl semimetal TaAs\cite{Weng2015PRX,Huang2015NC,Xu2015Science,Lv2015NP,Lv2015PRX,Yang2015NP}. The resultant asymmetry in the phonon line shape, conspicuous at low temperatures, diminishes continuously as the temperature increases. This anomalous behavior originates from the suppression of the electronic transitions near the Weyl points due to the decreasing occupation of electronic states below the Fermi level ($E_{F}$) with increasing temperature, as well as Pauli blocking caused by thermally excited electrons above $E_{F}$. Our findings not only elucidate the underlying mechanism governing the tunable Fano resonance, but also open a new route for exploring exotic physical phenomena through the properties of phonons in Weyl semimetals.}
\end{trivlist}
\unboldmath

%%%%%%%%%%%%%%%%%%%%%%%%%%%%%%%%%%%%%%%%%%%%%%%%%%%%%%%%%%%
%%
%%  Main body
%%
%%  Introduction
%

The Weyl semimetal (WSM) phase, a novel topological state of quantum matter, has been proposed to exist in materials with two non-degenerate bands crossing at $E_{F}$ in three-dimensional (3D) momentum space\cite{Wan2011PRB}. At the band crossing points (Weyl points), the electronic dispersion is linear in all three directions, resembling a 3D version of graphene, and the low-energy excitations can be described by Weyl equations, producing a condensed-matter realization of Weyl fermions\cite{Weyl1929ZP}. Recently, such a WSM state has been discovered in non-centrosymmetric transition-metal monoarsenides and monophosphides (TaAs, TaP, NbAs, and NbP)\cite{Weng2015PRX,Huang2015NC,Xu2015Science,Lv2015NP,Lv2015PRX,Yang2015NP}, where 12 pairs of Weyl points have been found. Since the Weyl points are located in close proximity to $E_{F}$ in these materials, interband electronic transitions near the Weyl points occur at a very low energy, 2$|\mu|$, where $|\mu|$ represents the chemical potential with respect to the Weyl points\cite{Ashby2014PRB,Xu2016PRB}. This energy scale overlaps optical phonon frequencies\cite{Liu2015PRB}. Consequently, strong coupling between electronic transitions near the Weyl points (Weyl fermions) and phonons may arise, enabling the study of Weyl-fermion physics through phonon behavior. Here, for the first time, we experimentally reveal strong coupling between Weyl fermions and phonons, as evidenced by a temperature-tunable Fano resonance in TaAs. These observations open a new avenue for exploring exotic quantum phenomena, such as the chiral anomaly\cite{Nielsen1983PLB,Hosur2013CRP,Huang2015PRX,Zhang2016NC}, in WSMs.

%%%%%%%%%%%%%%%%%%%%%%%%%%%%%%%%%%%%%%%%%%%%%%%%%%%%%%%%%%
%%  Experiment
%

High-quality single crystals of TaAs were synthesized through a chemical vapor transport method\cite{Huang2015PRX}. The as-grown crystals are polyhedrons with shiny facets up to 1.5~mm in size. X-ray diffraction (XRD) measurements reveal that the as-grown facets are the (001), (107) and (112) surfaces (Supplementary Information Fig.~S1). Systematic optical measurements were carried out on all three surfaces.

Figure~1a shows the far-infrared reflectivity $R(\omega)$ measured on the (107) surface of TaAs at 11 different temperatures from 5 to 300~K\cite{Xu2016PRB}. The relatively high $R(\omega)$ that approaches unity at zero frequency is consistent with the metallic nature of TaAs. A well-defined plasma edge in the far-infrared region suggests very low carrier density, in agreement with the tiny volumes enclosed by the Fermi surfaces (FSs) in this material\cite{Weng2015PRX,Huang2015NC,Lv2015NP,Xu2015Science}. In addition to the broad features in $R(\omega)$, a sharp feature can be clearly identified at $\sim$253~\icm\ as indicated by the arrow, which is associated with the infrared (IR)-active $A_{1}$ phonon mode\cite{Liu2015PRB} (Supplementary Information).

In order to gain direct information about this mode, we calculated the optical conductivity $\sigma_{1}(\omega)$ from $R(\omega)$ using a Kramers-Kronig analysis\cite{Xu2016PRB} (more details in Methods). Figure~1b displays $\sigma_{1}(\omega)$ of TaAs on the (107) surface in the far-infrared region at different temperatures. The low-frequency $\sigma_{1}(\omega)$ is dominated by a narrow Drude response alongside prominent linear features, whose origin and temperature dependence have been previously discussed in detail\cite{Xu2016PRB}. The $A_{1}$ mode manifests itself as a sharp peak in the $\sigma_{1}(\omega)$ spectrum, as indicated by the arrow. Figure~1c shows an enlarged view of $\sigma_{1}(\omega)$ in the frequency region of 240--270~\icm, where the $A_{1}$ mode can be seen more clearly. It is well known that in the absence of strong electron-phonon coupling, the phonon exhibits a symmetric line shape in $\sigma_{1}(\omega)$ that can be described by a Lorentz oscillator, as schematically illustrated in Fig.~1d. In contrast, strong electron-phonon coupling gives rise to an asymmetric phonon profile in $\sigma_{1}(\omega)$ that is referred to as the Fano line shape\cite{Fano1961PR,Tang2010NN,Li2012PRL} (Fig.~1e). As shown in Fig.~1c, the $A_{1}$ mode in TaAs exhibits a striking asymmetric line shape at low temperatures, which is an unequivocal signature of strong electron-phonon coupling. More interestingly, the asymmetry of the phonon line shape diminishes as the temperature rises, suggesting that the coupling-induced Fano resonance in TaAs can be tuned by temperature.

To quantify the temperature dependence of the $A_{1}$ mode, we extract the phonon line shape by subtracting a linear electronic background in a narrow frequency range at all measured temperatures, as shown in Fig.~2a. At each temperature, the phonon is fit with the Fano line shape\cite{Fano1961PR},
%%%%%%%%%%%%%%%%%%%%%%%%%%%
% Equation
%
\begin{equation}
\label{Fano}
\sigma_{1}(\omega)=\frac{2\pi}{Z_0} \frac{\Omega^2}{\gamma}
\frac{q^2 + \frac{4q(\omega-\omega_{0})}{\gamma} -1}{q^2 (1 + \frac{4(\omega-\omega_{0})^{2}}{\gamma^{2}})},
\end{equation}
where $Z_{0}$ is the vacuum impedance; $\omega_{0}$, $\gamma$, and $\Omega$ correspond to the resonance frequency, linewidth, and strength of the phonon, respectively; $q$ is a dimensionless parameter that describes the asymmetry of the Fano profile. A large $1/q^{2}$ indicates conspicuous asymmetry in the phonon line shape, while for $1/q^{2} = 0$, the symmetric Lorentz line shape is fully recovered. The solid lines in Fig.~2a represent the fitting curves, which describe the measured phonon line shapes reasonably well at all temperatures. This procedure also returns the temperature dependence of the fitting parameters.

Figure 2b depicts $1/q^{2}$ as a function of temperature. While $1/q^{2}$ adopts a large value of 1.1 at 5~K\cite{Kuzmenko2009PRL,Tang2010NN,Li2012PRL,Xu2015PRB}, it decreases dramatically with increasing temperature. This suggests that the $A_{1}$ mode is strongly coupled to a continuum of electron-hole excitations, and the resulting Fano resonance varies significantly with temperature. We first trace the origin of this coupling by examining the band structure of TaAs. Both first-principles calculations and ARPES measurements have revealed 12 pairs of Weyl points in TaAs\cite{Weng2015PRX,Huang2015NC,Xu2015Science,Lv2015NP}. These Weyl points are categorized into two types\cite{Weng2015PRX}. Four pairs in the $k_{z} = 0$ plane, about 2~meV above $E_{F}$, are defined as W1, while another eight pairs off the $k_{z} = 0$ plane, lying about 21~meV below $E_{F}$, are named W2. Interband electronic transitions in the vicinity of a Weyl point start at $\omega =$ 2$|\mu|$\cite{Ashby2014PRB,Xu2016PRB}, making it easy to calculate that electronic transitions near W2 turn on at $\omega =$ 42~meV ($\sim$336~\icm). Thus electronic transitions at the frequency of the $A_{1}$ mode (253~\icm) do not occur near W2. This implies that the $A_{1}$ mode is unlikely to be coupled to the electronic transitions near W2. However, electronic transitions near W1 set in at $\omega >$ 4~meV ($\sim$32~\icm), and can therefore overlap the frequency of the $A_{1}$ mode, suggesting that this mode is coupled to the electronic transitions near W1.

Having attributed the asymmetric line shape of the $A_{1}$ mode to its strong coupling with the electronic transitions near W1, we proceed to understand the temperature dependence of this mode. In WSMs, since the Weyl points are in close proximity to $E_{F}$, electronic transitions near the Weyl points can be dramatically affected by thermal excitations, thus changing the line shape of the phonon that is coupled to these transitions. Figure~3 shows the band structure along three momentum directions near W1 in TaAs. The occupation probabilities of the electronic states in these bands (color maps) are calculated using the Fermi-Dirac distribution function at three different temperatures (5, 150 and 300~K), from which we see that the occupation of the electronic states near W1 depends strongly on the temperature. At 5~K (Fig.~3a-c), the electronic states below $E_{F}$ are fully occupied (blue), while the states above $E_{F}$ are empty (white). In this case, interband transitions at the energy of the $A_{1}$ mode $\hbar\omega_{0}$ are strong, as indicated by the thick arrows. As the temperature increases through 150 to 300~K, thermal excitations cause vacant states to appear below $E_{F}$ and electronic states above $E_{F}$ to be partially occupied. Electronic transitions at $\hbar\omega_{0}$ are significantly suppressed (illustrated by the thin arrows in Fig.~3), because the available initial states for these transitions decrease, and many of the final states are Pauli blocked by thermally excited electrons. This suppression of the electronic transitions near W1 is directly responsible for the change in the $A_{1}$ phonon line shape.

For a quantitative analysis, the dimensionless parameter $q$ in the Fano theory is given by\cite{Fano1961PR,Tang2010NN}
%%%%%%%%%%%%%%%%%%%%%%%%%%%
% Equation
%
\begin{equation}
\label{qTD}
q = \frac{1}{\pi V_{e-ph} D_{e-h}(\omega_{0}, T)} \times \frac{\mu_{ph}}{\mu_{e-h}},
\end{equation}
where $V_{e-ph}$ is the electron-phonon coupling strength; $D_{e-h}(\omega_{0}, T)$ is the joint electron-hole pair density of states at the frequency of the $A_{1}$ mode $\omega_{0}$ for a given temperature $T$; $\mu_{ph}$ and $\mu_{e-h}$ represent the optical matrix elements for phonon and electron-hole pair excitations, respectively. In this equation, we note that raising $T$ mainly modifies $D_{e-h}(\omega_{0}, T)$ by thermally exciting electrons to the electronic states above $E_{F}$ and creating holes below $E_{F}$. Near the Weyl points, the finite-temperature joint electron-hole pair density of states at $\hbar\omega_{0}$ takes the form
%%%%%%%%%%%%%%%%%%%%%%%%%%%
% Equation
%
\begin{equation}
\label{JDOS}
D_{e-h}(\omega_{0}, T) = D_{e-h}^{0}(\omega_{0})\left[f(-\frac{\hbar \omega_{0}}{2})-f(\frac{\hbar \omega_{0}}{2})\right],
\end{equation}
where $f(\epsilon) = 1/[e^{(\epsilon-\mu)/k_{B}T}+1]$ is the Fermi function, with $\epsilon$ representing the energy of the single-particle state with respect to the Weyl points and $D_{e-h}^{0}(\omega_{0})$ the zero-temperature joint electron-hole pair density of states at $\hbar\omega_{0}$. The red solid curve in Fig.~2b is the least-squares fit to the experimental temperature dependence of $1/q^{2}$ (blue solid circles) using Eq.~(\ref{qTD}). The excellent agreement between our experimental data and the model further underlines the intimate link between the line shape of the $A_{1}$ mode and the electronic transitions near W1, which is continuously suppressed with increasing temperature due to the reduced occupation of the electronic states below $E_{F}$ and Pauli-blocking from thermally excited electrons above $E_{F}$.

A careful examination of the temperature dependence of the $A_{1}$ phonon linewidth $\gamma$ (Fig.~2c) leads us to the same conclusion. In the case of weak electron-phonon coupling, phonon decay is dominated by the anharmonic effect: a zone-center phonon decays into two acoustic modes with the same frequencies and opposite momenta\cite{Klemens1966PR,Menendez1984PRB}. The temperature dependence of the phonon linewidth $\gamma^{ph-ph}(T)$ for this process follows
%%%%%%%%%%%%%%%%%%%%%%%%%%%
% Equation
%
\begin{equation}
\label{anharmonic}
\gamma^{ph-ph}(T) = \gamma^{ph-ph}_{0}\left(1+\frac{2}{e^{\frac{\hbar\omega_{0}}{2k_{B}T}}-1}\right),
\end{equation}
where $\gamma^{ph-ph}_{0}$ is the residual linewidth at zero temperature. Apparently, this model does not account for the behavior of the $A_{1}$ mode in TaAs, since it gives an increasing $\gamma$ as the temperature is raised, which is opposite to our experimental result. Instead, strong electron-phonon coupling must be taken into account to understand the temperature dependence of $\gamma$ in TaAs. In a system with strong electron-phonon coupling, a phonon can also decay by creating an electron-hole pair\cite{Bonini2007PRL}. This process is sensitive to $D_{e-h}(\omega_{0}, T)$ and is thus suppressed with increasing temperature due to thermal excitations, resulting in a temperature-dependent phonon linewidth $\gamma^{e-ph}(T)$
%%%%%%%%%%%%%%%%%%%%%%%%%%%
% Equation
%
\begin{equation}
\label{ephdecay}
\gamma^{e-ph}(T) = \gamma^{e-ph}_{0}\left[f(-\frac{\hbar \omega_{0}}{2})-f(\frac{\hbar \omega_{0}}{2})\right],
\end{equation}
where $\gamma^{e-ph}_{0}$ represents a residual linewidth. Consequently, the temperature-dependent linewidth of a phonon mode that is strongly coupled to electronic excitations is given by $\gamma(T) = \gamma^{ph-ph}(T) + \gamma^{e-ph}(T)$. This equation gives an excellent description to the measured temperature dependence of the linewidth for the $A_{1}$ mode in TaAs, as shown by the red solid curve in Fig.~2c. The above observations explicitly demonstrate that the $A_{1}$ mode is strongly coupled to the electronic transitions near W1, and both the line shape and linewidth of this phonon are closely tied to these transitions, which can be continuously tuned by temperature through varying the occupation of the electronic states near W1. It is worth pointing out that changing the occupation of the electronic states near W1 via other methods should also induce a noticeable change in the line shape of the $A_{1}$ mode in TaAs.

Finally, to lend further credence to our experimental results and analysis, we performed optical measurements at 12 different temperatures on the (112) surface. Essentially identical behavior was revealed for the $A_{1}$ mode (Supplementary Information Fig.~S3). We then utilized the same methods and models to analyze the line shape and linewidth of this mode observed on the (112) surface (Supplementary Information Fig.~S4), reaching the same conclusions.

%%%%%%%%%%%%%%%%%%%%%%%%%%%%%%%%%%%%%%%%%%%%%%%%%%%%%%%%%%%%%%%%
%
%% Discussion
%

The observed strong coupling between Weyl fermions and phonons enables the study of exotic quantum phenomena in WSMs by tracking the properties of phonons. One interesting proposal derived from our findings is to reveal experimental evidence for the chiral anomaly\cite{Nielsen1983PLB,Hosur2013CRP,Huang2015PRX,Zhang2016NC}, in which the application of parallel electric ($\mathbf{E}$) and magnetic ($\mathbf{B}$) fields pumps electrons from one Weyl point to the other with opposite chirality at a rate proportional to $\mathbf{E}\cdot\mathbf{B}$, leading to a shift of $E_{F}$ in opposite directions at different Weyl points. This $E_{F}$ shift caused by the chiral anomaly can significantly affect the electronic transitions near the Weyl points. In TaAs, we have demonstrated that the $A_{1}$ mode is strongly coupled to the electronic transitions near W1, and its line shape is sensitive to the amplitude of these transitions. These facts suggest that tracking the $A_{1}$ phonon line shape in TaAs as a function of the amplitudes of $\mathbf{E}$ and $\mathbf{B}$, as well as the angle between them, is likely to provide convincing evidence for the chiral anomaly.

In conclusion, we have found compelling evidence for strong coupling between the IR-active $A_{1}$ phonon mode and electronic transitions near W1 (Weyl fermions) in TaAs through the observation of a Fano resonance. Varying the temperature, which changes the occupation of the electronic states in proximity to $E_{F}$, can continuously tune the amplitude of the electronic transitions near W1, thus manipulating the line shape of the $A_{1}$ mode. Our findings suggest a new approach for investigating exotic quantum phenomena in WSMs, such as the chiral anomaly, through the behavior of phonons.

{\footnotesize
\subsubsection*{Methods}

\begin{trivlist}
\item[]{\bf Sample synthesis.}\hspace{2mm} High-quality TaAs single crystals were grown through a chemical vapor transport method~\cite{Huang2015PRX}. A previously reacted polycrystalline TaAs was filled in a quartz ampoule using iodine (2~mg/cm$^{3}$) as the transporting agent. After evacuating and sealing, the ampoule was kept at the growth temperature for three weeks. Large polyhedral crystals with dimensions up to 1.5~mm are obtained in a temperature field of $\Delta T$ = 1150--1000~$^{\circ}$C.
\item[]{\bf Optical measurements.}\hspace{2mm} The frequency-dependent reflectivity $R(\omega)$ was measured at a near-normal angle of incidence on a Bruker VERTEX 80v FTIR spectrometer. In order to accurately measure the absolute $R(\omega)$ of the samples, an \emph{in situ} gold overcoating technique~\cite{Homes1993} was used. Data from 40 to 15\,000~\icm\ were collected at 11 different temperatures from 300 down to 5~K on three different surfaces of as-grown single crystals in an ARS-Helitran cryostat. Since a Kramers-Kronig analysis requires a broad spectral range, $R(\omega)$ was extended to the UV range (up to $50\,000$~\icm) at room temperature with an AvaSpec-2048 $\times$ 14 fiber optic spectrometer.
\item[]{\bf Kramers-Kronig analysis.}\hspace{2mm} The real part of the optical conductivity $\sigma_{1}(\omega)$ has been determined via a Kramers-Kronig analysis of $R(\omega)$~\cite{Dressel2002}. Below the lowest measured frequency, we used a Hagen-Rubens ($R = 1 - A\sqrt{\omega}$) form for the low-frequency extrapolation. Above the highest measured frequency, we assumed a constant reflectivity up to 12.5~eV, followed by a free-electron ($\omega^{-4}$) response.
\end{trivlist}
}

%%%%%%%%%%%%%%%%%%%%%%%%%%%%%%%%%%%%%%%%%%%%%%%%%%%%%%%%%%%%%%%%%%%%%%%%%%
%%
%% The bibliography (BibTeX)
%%

%% Put the bibliography here. BibTex bibliography files cannot be accepted.
%% LaTeX submission must contain all references within the manuscript .tex file itself.
%% Most people will use BiBTeX in which case the environment below should be replaced with the \bibliography{} command.

%%%%%%%%%%%%%%%%%%%%%%%%%%%%%%%%%%%%%%%%%%%%%%%%%%%%%%%%%%%%%%%%%%%%%%%%%%%%%%%%%%%%%
%%
%%   Acknowledgements, Author contributions, Competing financial interests ...
%%

{\footnotesize
\subsubsection*{Acknowledgements}
The authors acknowledge Ricardo Lobo, Yongkang Luo, Simin Nie and Hongming Weng for illuminating discussions. Work at IOP CAS was supported by MOST (973 Projects No. 2015CB921303, No. 2015CB921102, No. 2012CB921302, and No. 2012CB821403), and NSFC (Grants No. 91121004, No. 91421304, and No. 11374345). Work at LANL was performed at the Center for Integrated Nanotechnologies, a US Department of Energy, Office of Basic Energy Sciences user facility, and funded by the LANL LDRD program and by the UC Office of the President under the UC Lab Fees Research Program, Grant ID No. 237789. H. Xiao is supported by NSFC, Grant No. U1530402.

\subsubsection*{Author contributions}
B.X. carried out the optical measurements with the assistance of K.W., R.Y., W.Z. and J.Y.L.; L.X.Z. and G.F.C. synthesized the single crystals; B.X., Y.M.D., H.X., A.J.T., D.A.Y., R.P.P. and X.G.Q. analyzed the data; S.A.T. and J.X.Z. contributed to theoretical models; Y.M.D. wrote the manuscript; all authors made comments on the manuscript; R.P.P. and X.G.Q. supervised the project.

\subsubsection*{Additional information}
Supplementary information is available in the online version of the paper. Reprints and permissions information is available online. Correspondence and requests for materials should be addressed to R.P.P. and X.G.Q.

\subsubsection*{Competing financial interests}
The authors declare no competing financial interests.
}

%%%%%%%%%%%%%%%%%%%%%%%%%%%%%%%%%%%%%%%%%%%%%%%%%%%%%%%%%%%%%%%%%%%%%%%%%
%%
%% Figure captions
%%
%% Nature series want the figures to be submitted separately.
%% \includegraphics will be ignored in the whole .tex file.
%% Before online submission, make sure to remove all the \includegraphics.
%%

%%%%%%%%%%%%%%%%%%%%%%%%%%%%%%%%%%%%%%%%%%%%%%%%%
%
%  Figure 1
%
\clearpage
\centerline{\includegraphics[width=0.95\columnwidth]{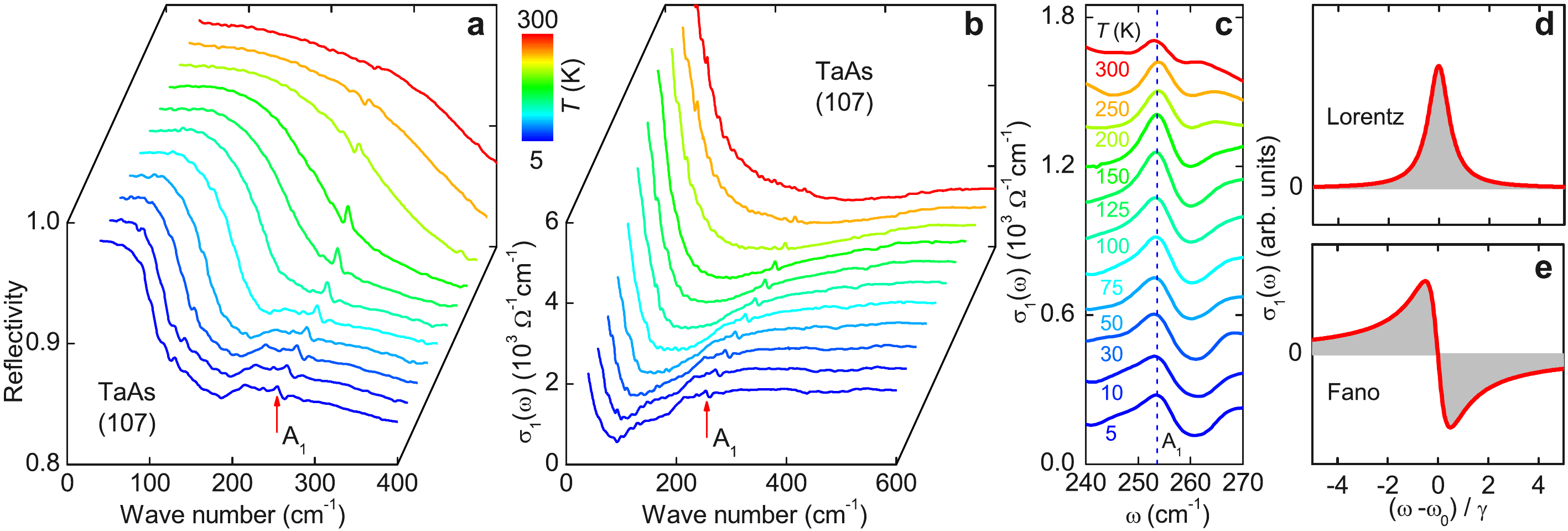}}%
\vspace*{0.8cm}
\noindent{\textsf{\textbf{Figure 1 $|$ Reflectivity and optical conductivity of TaAs.} \textbf{a}, Reflectivity of TaAs in the far-infrared region measured at different temperatures on the (107) surface. \textbf{b}, Optical conductivity of TaAs on the (107) surface up to 600~\icm\ at different temperatures. \textbf{c}, Enlarged view of the optical conductivity in the region of the infrared-active $A_{1}$ mode at $\sim$253~\icm. \textbf{d}, Schematic of the symmetric Lorentz oscillator, which describes the phonon line shape in the optical conductivity without strong electron-phonon coupling. \textbf{e}, Schematic of the asymmetric Fano resonance, used to describe the phonon line shape in the presence of strong electron-phonon coupling.}}
%
%%%%%%%%%%%%%%%%%%%%%%%%%%%%%%%%%%%%%%%%%%%%%%%%%
%
%  Figure 2
%
\clearpage
\centerline{\includegraphics[width=0.9\columnwidth]{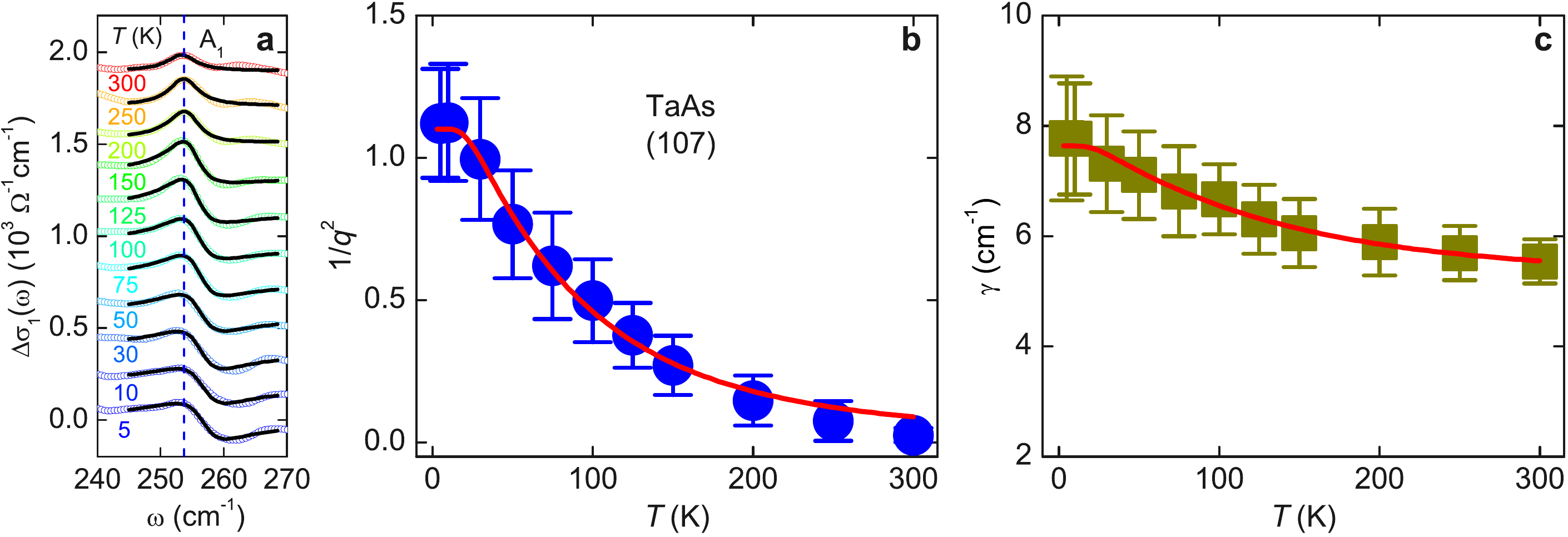}}%
\vspace*{0.8cm}
\noindent{\textsf{\textbf{Figure 2 $|$ Fano fit to the phonon line shape and temperature dependence of the fitting parameters for TaAs.} \textbf{a}, Line shape of the $A_{1}$ phonon, with the electronic background subtracted at different temperatures. The black solid lines through the data denote the Fano fitting results. \textbf{b} and \textbf{c}, Temperature dependence of the Fano parameter $1/q^{2}$ and the line width $\gamma$ of the $A_{1}$ mode, respectively. The red solid lines through the data in each panel represent the modeling results.}}

%%%%%%%%%%%%%%%%%%%%%%%%%%%%%%%%%%%%%%%%%%%%%%%%%
%
%  Figure 3
%
\clearpage
\centerline{\includegraphics[width=0.7\columnwidth]{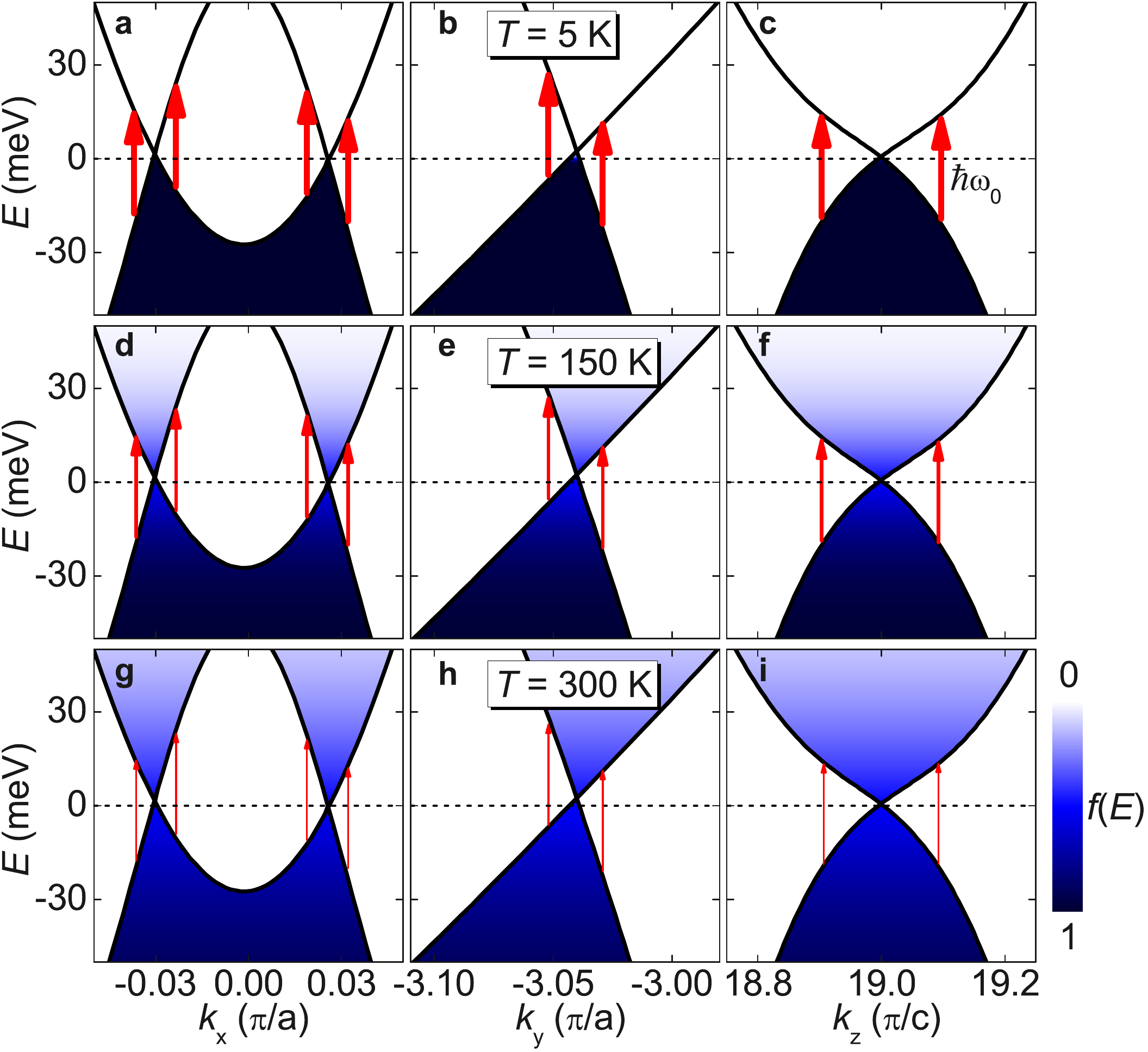}}%
\vspace*{0.8cm}
\noindent{\textsf{\textbf{Figure 3 $|$ Occupation probability of the electronic states near the Weyl points in TaAs at different temperatures.} Band structure along three different momentum directions near the Weyl points W1 in TaAs. The color maps, which are calculated from the Fermi-Dirac distribution function, denote the occupation probability of the electronic states at different temperatures of 5~K (\textbf{a}--\textbf{c}), 150~K (\textbf{d}--\textbf{f}) and 300~K (\textbf{g}--\textbf{i}). The red arrows represent the electronic transitions at the energy of the $A_{1}$ mode $\hbar\omega_{0}$. The thickness of each arrow schematically depicts the transition amplitude.}}

%%%%%%%%%%%%%%%%%%%%%%%%%%%%%%%%%%%%%%%%%%%%%%%%%%%%%%%%%%%%%%%%%%%%%%%%
%% TABLES
%%
%% If there are any tables, put them here. Please submit tables at the end of your text document.
%%

\end{document}